\documentclass[prl,letterpaper,aps,10pt,superscriptaddress,twocolumn,floatfix,showpacs]{revtex4-1}
\usepackage{amsmath,amsfonts,dsfont,graphicx,amssymb,microtype,braket,xcolor,upgreek,mathtools,physics}
\usepackage{placeins}
\usepackage[colorlinks, linkcolor=blue, citecolor=blue, urlcolor=blue, breaklinks=true]{hyperref}
\DeclareGraphicsExtensions{.pdf}

\binoppenalty=\maxdimen
\relpenalty=\maxdimen

\newcommand{\ts}[1]{_\text{#1}}
\newcommand{\me}{\mathrm{e}}
\newcommand{\mi}{\mathrm{i}}
\newcommand{\wt}{\widetilde}
\newcommand{\bk}{\mathbf{k}}
\newcommand{\bq}{\mathbf{q}}

\newcommand{\bE}{\mathbf{E}}
\newcommand{\br}{\mathbf{r}}

\newcommand{\calA}{\mathcal{A}}
\newcommand{\calM}{\mathcal{M}}
\newcommand{\calH}{\mathcal{H}}

\newcommand{\calN}{\mathcal{N}}

\newcommand{\gaz}{\widetilde\Gamma(0)}
\newcommand{\iu}{\mathrm{i}}
\newcommand{\diff}{\mathrm{d}}

\providecommand{\ave}[1]{\langle#1\rangle}
\providecommand{\expe}[1]{\left\langle#1\right\rangle}
\providecommand{\mean}[3]{\langle#1|#2|#3\rangle} 
\newcommand{\tR}{{t_{\text{RT}}}}

\newcommand{\di}{\textrm{d}}

\newcommand{\ha}{a}

\usepackage{bm}
\usepackage{booktabs}

\begin{document}

\newcommand{\MPL}{Max  Planck  Institute  for  the  Science  of  Light,  D-91058  Erlangen,  Germany}
\newcommand{\FAU}{Department of Physics, Friedrich-Alexander-Universit\"{a}t Erlangen-N\"urnberg (FAU), D-91058 Erlangen, Germany}
\newcommand{\UCSD}{Department of Chemistry and Biochemistry, University of California San Diego, La Jolla, California 92093, USA}
\title{Metasurface-based hybrid optical cavities for chiral sensing}
\author{Nico S. Ba\ss ler}
\affiliation{\MPL}
\affiliation{\FAU}
\author{Andrea Aiello}
\affiliation{\MPL}
\author{Kai P. Schmidt}
\affiliation{\FAU}
\author{Claudiu Genes}
\affiliation{\MPL}
\affiliation{\FAU}
\author{Michael Reitz}
\affiliation{\MPL}
\affiliation{\UCSD}
\date{\today}

\begin{abstract}
Quantum metasurfaces, i.e., two-dimensional subwavelength arrays of quantum emitters, can be employed as mirrors towards the design of hybrid cavities, where the optical response is given by the interplay of a cavity-confined field and the surface modes supported by the arrays. We show that, under external magnetic field control, stacked layers of quantum metasurfaces can serve as helicity-preserving cavities. These structures exhibit ultranarrow resonances and can enhance the intensity of the incoming field by orders of magnitude, while simultaneously preserving the handedness of the field circulating inside the resonator, as opposed to conventional cavities. The rapid phase shift in the cavity transmission around the resonance can be exploited for the sensitive detection of chiral scatterers passing through the cavity. We discuss possible applications of these resonators as sensors for the discrimination of chiral molecules.
\end{abstract}


\maketitle
Conventional isotropic (e.g., metallic) mirrors reverse the handedness (or helicity) of circularly polarized light by turning right-circularly polarized (RCP) light into left-circularly polarized (LCP) light and vice versa \cite{coles2012chirality, barnett2012duplex}. This makes it impossible to realize helicity-preserving (HP) cavities or even chiral cavities (i.e., cavities only supporting light modes of a certain handedness), with conventional mirrors \cite{plum2015chiral, aiello2022helicity}. There is however a great current scientific and technological interest in the design of HP mirrors and resonators \cite{hentschel2017chiral, semnani2020spin, feis2020helicity, voronin2022single}, in particular for the enhancement of so-called dichroic effects. Dichroism refers to the (typically weak) differential absorption of circularly polarized light by chiral scatterers such as molecular enantiomers \cite{tang2010optical}. Enhancing dichroic effects with optical resonators can result in better sensitivities for the discrimination of molecular enantiomers \cite{scott2020on, mohammadi2018nanophotonic, genet2022chiral, mauro2023chiral}, a desired task for biochemical applications. In the strong light-matter coupling regime, chiral cavities have furthermore been proposed to create novel light-dressed states of matter by breaking the time-reversal symmetry in materials, leading to the emerging field of chiral polaritonics~\cite{huebener2021strong, schaefer2023chiral}. \newline\indent In this work we show that HP cavities can be implemented with quantum metasurfaces employed as mirrors. These structures have emerged as platforms for achieving strong and highly directional light-matter interactions and can most prominently be realized with cold atoms trapped in optical lattices~\cite{rui2020asubradiant}. They can exhibit close to perfect reflection of incoming light~\cite{bettles2016enhanced, Shahmoon2017,rui2020asubradiant, ballantine2020optical, alaee2020quantum, ballantine2021cooperative} and have numerous other applications e.g., as platforms for topological quantum optics~\cite{bettles2017topological,Perczel2017,Perczel2017_2}, nonlinear quantum optics~\cite{bettles2020quantum, parmee2021bistable, morenocardoner2021quantum, rusconi2021exploiting, srakaew2023subwavelength, pedersen2023quantum} or quantum information processing~\cite{plankensteiner2015selective, facchinetti2016storing, manzoni2018optimization,grankin2018free, guimond2019subradiant, Bekenstein2020}. The main ingredient of our approach is to manipulate the polarization of the incoming light field via the orientation of the effective two-level systems that make up the metasurfaces, which can for instance be tuned via an external magnetic field. \newline\indent More generally, this work falls within the scope of \textit{hybrid cavities}, i.e., the design of optical resonators going beyond the simple textbook picture of a single electromagnetic mode confined between two non-reactive mirrors. Instead, strongly dispersive optical elements such as photonic crystals or plasmonic metasurfaces are used as reflectors \cite{Zhou2014, cernotik2019cavity, denning2019quantum, fitzgerald2021cavity, binalam2021ultra} with the aim to surpass the performance of standard cavities, implying a highly non-Markovian behavior of the cavity as characterized by non-Lorentzian, typically Fano-type lineshapes \cite{Miroshnichenko2010}. \\

\begin{figure*}[t]
  \centering
  \includegraphics[width=1.0\textwidth]{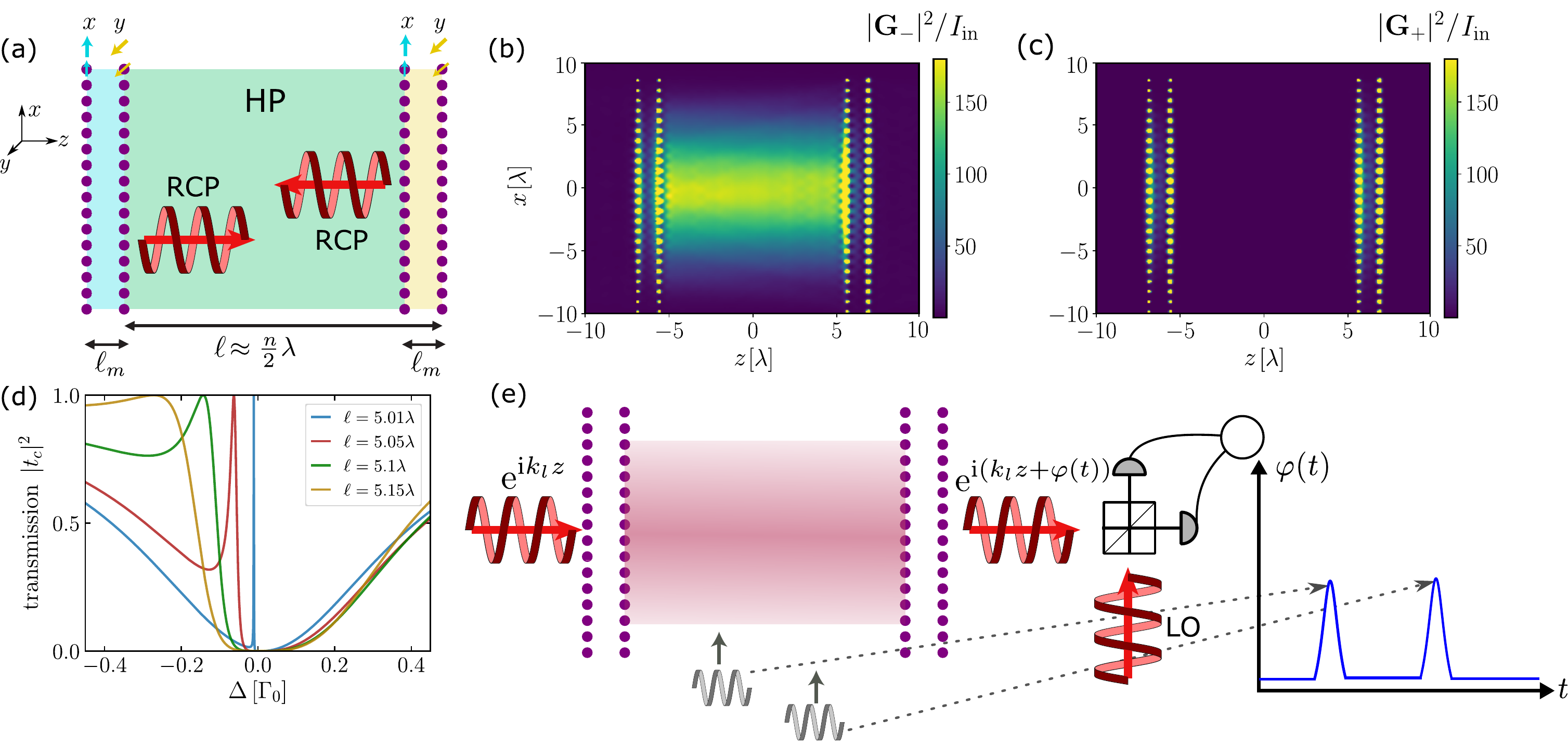}
  \caption{\textit{Helicity-preserving metasurface cavity.} (a) A HP quantum metasurface cavity of length $\ell\approx n\lambda/2$ can be constructed with composite mirrors of orthogonal dipole orientation (e.g., along $x$ and $y$), with an appropriate relative phase separation of $\phi_m=k_l\ell_m=\pi/2+2\pi n_m$. (b), (c) Absolute value squared of RS vectors $\mathbf G_\pm (\mathbf R)$ for a cavity length $\ell=12.505\lambda$, $\ell_m=5\lambda/4$, for square lattices with lattice spacing $a=0.8\lambda$ illuminated by RCP light with input polarization vector $\bE\ts{in}^{(+)}=(1,\iu,0)^\top/\sqrt{2}$ close to the cavity resonance. The plots are normalized to the input intensity $I\ts{in}=|\bE\ts{in}^{(+)}|^2$. The metasurfaces are defined with a finite curvature radius, leading to Gaussian confinement of the mode (beam waist $w_0=8\lambda$). (d)  Cavity transmission $|t_c|^2$ as a function of the laser detuning $\Delta$ for different cavity lengths $\ell$.  (e) Chiral sensing with HP metasurface cavities: Ideal chiral scatterers passing through a cavity with the same handedness yield a phase shift in the cavity transmission output which can for instance be measured by homodyne detection (LO: local oscillator) and leads to a phase detection sequence as schematically illustrated in the diagram.}
  \label{fig:fig1}
\end{figure*}

\noindent \textbf{HP mirror} -- Let us present an implementation procedure for a HP mirror using a stacked system of quantum metasurfaces. We start by introducing the formalism for a single metasurface~\cite{bassler2023linear}. To this end, we consider a 2D quasi-infinite quantum emitter array where the emitters are situated in the $xy$ plane at positions $\br_j$. For simplicity, one may imagine a square lattice, however most of the results derived in the following are equally valid for other Bravais lattices and can furthermore also be extended to non-Bravais lattices \cite{Perczel2017, bassler2023linear}. The layer is comprised of $\cal N$ emitters with internal electronic structure described by a $J=0\to J=1$ transition at transition frequency $\omega_0$. In the following, we will  work in the Cartesian polarization basis. The transition dipole operator for each emitter can be written as $\mathbf{d}=\sum_\nu \mathbf{d}_{\nu}\sigma_\nu+\mathrm{h.c.}$ with $\mathbf{d}_{\nu}=\bra{g}\mathbf{d}\ket{\nu}$ and $\sigma_\nu=\ket{g}\bra{\nu}$ is the corresponding lowering operator for each  electronic transition ($\nu=x,y,z$). In addition, we consider a laser drive entering from the left in the form of a plane wave with positive-frequency amplitude $\bE\ts{in}^{(+)}=(E_{\text{in},x}, E_{\text{in},y},0)^\top$ and laser frequency $\omega_l=2\pi c/\lambda=ck_l$ where $\lambda$ and $k_l$ are the laser wavelength and wavenumber, respectively.

In a frame rotating at the laser frequency, the Hamiltonian describing the dynamics of the emitter array is given by the sum of the free evolution and the dipole-dipole interaction ($\hbar = 1$)
\begin{align}
  \label{eq:freeham}
  \calH_0+\calH\ts{d-d}\!=-\Delta\sum_{j,\nu}\sigma_{j,\nu}^\dagger\sigma_{j,\nu}^{\phantom{\dagger}}\!+\!\sum_{j,j',\nu,\nu'}\Omega_{jj'}^{\nu\nu'}\sigma_{j,\nu}^\dagger\sigma_{j',\nu'}^{\phantom{\dagger}},
\end{align}
with the laser detuning $\Delta=\omega_l-\omega_0$ and $\sigma_{j,\nu}$ is the lowering operator for the $\nu$-transition within a particular emitter $j$. Assuming normally incident illumination, the laser drive adds as $\calH_l=\sum_{j,\nu}(\eta_\nu^{\phantom{\dagger}}\sigma_{j,\nu}^\dagger+\mathrm{h.c.})$ with Rabi frequencies $\eta_{\nu}^{\phantom{i}}=d_{\nu}^{\phantom{i}} E^{(+)}_{\mathrm{in}, {\nu}}$ and $\eta_z=0$. In addition to the coherent processes, the collective loss of excitations due to spontaneous emission is described by the Lindblad term
\begin{align}
  \label{eq:lindblad}
  \mathcal L [\rho]\!=\!\sum_{j,j',\nu,\nu'}\Gamma_{jj'}^{\nu\nu'}\left[\sigma_{j,\nu}^{\phantom{\dagger}}\rho \sigma_{j',\nu'}^\dagger\!-\!\frac{1}{2}\left\{\sigma_{j,\nu}^\dagger\sigma^{\phantom{\dagger}}_{j',\nu'},\rho\right\}\right],
\end{align}
where the last term denotes an anticommutator and the diagonal elements describe the independent spontaneous emission of the emitters $\Gamma_{jj}^{\nu \nu'}=\Gamma_0\delta_{\nu\nu'}$ with $\Gamma_0=\omega_0^3 d^2/(3\pi\epsilon_0 c^3)$ (we assume the dipole moments to be identical in the following $d_\nu\equiv d$). The rates $\Omega_{jj'}^{\nu\nu'}$, $\Gamma_{jj'}^{\nu\nu'}$ describe coherent/incoherent scattering of photons between emitters $j$ and $j'$ and between transitions $\nu$ and $\nu'$ and can be derived as real and imaginary parts of the photonic Green's tensor (see App.~\ref{sec:AppendixA}) as
\begin{align}
  \Omega_{jj'}^{\nu\nu'}-\mi\frac{\Gamma_{jj'}^{\nu\nu'}}{2}=-\mu_0 \omega_0^2 \,\mathbf{d}_\nu^*\cdot \mathbf{G}(\mathbf{r}_{jj'}) \cdot\mathbf{d}_{\nu'}^{\phantom{*}},
\end{align}
expressed in terms of the vacuum permeability $\mu_0$ and depending on the interparticle separation $\mathbf{r}_{jj'}=\mathbf r_{j'}-\mathbf r_j$. The Green's tensor is defined such that the real part of the self-interaction at $j=j'$ vanishes. From the steady-state solution of the quantum master equation $\dot\rho=\mi[\rho, \calH_0\!+\!\calH\ts{d-d}\!+\!\calH_l]\!+\!\mathcal{L}[\rho]$, the dipole amplitudes and thereby the transmitted and reflected fields can be computed (see App.~\ref{sec:AppendixA}). The transmission matrix of the metasurface connecting the polarization components of the input field to the outgoing field expresses as \cite{bassler2023linear}
\begin{align}
  \boldsymbol{\mathcal{T}}_m=\mathds{1}+\mi\frac{\widetilde\Gamma (0)}{2d^2}\boldsymbol{\alpha}\ts{red},
\end{align}
where $\wt\Gamma (0)$ is the effective decay rate at zero quasi-momentum and $\boldsymbol{\alpha}\ts{red}$ is the 2D polarizability tensor of the metasurface, relating the induced dipole moment to the incoming electric field. However, in the limit of large external magnetic fields $\mu\abs{\mathbf B}\gg\wt\Gamma (0)$ (magnetic moment $\mu$), all dipole transitions orthogonal to the magnetic field direction become very off-resonant and one may focus on the polarization component in the direction of the magnetic field, thereby reducing the description to an effective two-level model \cite{bassler2023linear}. The transmission amplitude of the metasurface for a single component $E_{\text{in},\nu}$ is then simply given by~\cite{bettles2016enhanced, Shahmoon2017}
\begin{align}
  t_m=1+\frac{\mi\wt\Gamma (0)/2}{\wt\Omega(0) -\Delta-\mi\gaz/2},
\end{align}
where $\wt\Omega (0)=\sum_j \Omega_{0j}^{\nu\nu}$, $\wt\Gamma (0)=\sum_j \Gamma_{0j}^{\nu\nu}$ describe the dipole-induced collective frequency shift and decay rate arising from the $\nu$ transition dipoles (for an arbitrary index $0$ on the array). The complex transmission and reflection amplitudes are connected as $t_m=1+r_m$ while $|t_m|^2+|r_m|^2=1$. Most notably, if the laser frequency matches the collective metasurface resonance $\omega_l=\omega_\nu+\wt \Omega (0)$, perfect reflection of incoming light is obtained as $|r_m|^2=1$. In the following, for the sake of clarity, we proceed with the simplified two-level description. Finally, to obtain a HP mirror, we consider now two copies of quantum metasurfaces separated by a distance $\ell_m$, one with dipoles pointing in $x$-direction and one with dipoles pointing in $y$-direction with a path length difference of $k_l\ell_m=\phi_m=2\pi n_m+\pi/2$ ($n_m\in\mathbb{N}_0$) between the two polarizations. The combination of these two mirrors is a helicity-perserving mirror. The path length difference rotates the $y$-polarization components by $\pi$, thereby reversing the mirror operation which does not conserve helicity for an ordinary mirror. A full transfer matrix calculation showing this can be found in App.~\ref{sec:AppendixE}.\\

\noindent \textbf{HP cavity} -- A HP optical cavity can now be simply implemented by two HP metasurface mirrors separated by a distance $\ell$ (see Fig.~\ref{fig:fig1}(a)). The two layers making up the mirror consist of dipoles with perpendicular dipole orientations, leading to vanishing interactions between the two cavities in the far field. For simplicity, we thus continue the discussion for a single cavity while keeping in mind that the actual setup consists of two noninteracting copies. A full discussion for both polarization components can be found in App.~\ref{sec:AppendixE}. Solving the coupled-dipole equations and neglecting the contributions from all evanescent terms, a simple expression for the total transmitted field can be obtained as $E^{(+)} (z>\ell)=t_c E^{(+)}\ts{in}\me^{\mi k_l z}$ with the cavity transmission coefficient (assuming $k_0\approx k_l$, for derivation see App.~\ref{sec:AppendixC})
\begin{align}
  \label{eq:dipoletheory}
  t_c=\frac{\left(\Delta-\wt\Omega (0)\right)^2}{\left(\Delta-\wt\Omega (0)+\mi\frac{\gaz}{2}\right)^2+\frac{\gaz^2}{4}\me^{2\mi k_l \ell}}.
\end{align}
We remark that instead of solving the coupled-dipole equations for the two arrays, the same result can be obtained from classical transfer matrix theory \cite{cernotik2019cavity, reitz2022cooperative, pedersen2023quantum} where the transfer matrix of a single metasurface can be expressed in terms of the mirror polarizability $\zeta_m=-\mi r_m/t_m=\wt\Gamma (0)/[2(\wt\Omega (0)-\Delta)]$  as
\begin{align}\mathbf T_m=
  \begin{pmatrix}
    1+\mi\zeta_m & \mi\zeta_m \\
    -\mi\zeta_m & 1-\mi\zeta_m
  \end{pmatrix}.
\end{align}
The total transfer matrix is  then simply obtained as $\mathbf T=\mathbf T_m \mathbf T_f \mathbf T_m$ with the free space propagation matrix $\mathbf T_f = \mathrm{diag}(\me^{\mi k_l \ell}, \me^{-\mi k_l \ell})$. The condition that the transmission ought to equal unity at the cavity resonance $|t_c|^2=1$, yields the following expression for the cavity resonance
\begin{align}
  \label{eq:resonance}
  \Delta-\wt\Omega (0) =-\frac{\wt\Gamma (0)}{2}\tan (k_l\ell).
\end{align}
To demonstrate that the resulting cavity consisting of two HP mirrors indeed conserves the helicity, we compute the Riemann-Silberstein (RS) vectors \cite{birula2013silberstein}
\begin{equation}
  \label{eq:silberstein_vectors}
  \mathbf G_{\pm}(\mathbf R)=\frac{1}{\sqrt{2}}\left(\mathbf E (\mathbf R)\pm\iu \mathcal Z\mathbf H (\mathbf R)\right),
\end{equation}
which describe the combined electromagnetic field of chiral polarization and $\mathcal Z=(\epsilon_0c)^{-1}$ is the vacuum impedance. The absolute value of these quantities is plotted in Figs.~\ref{fig:fig1}(b), (c), for RCP light entering the cavity, confirming that the cavity preserves the helicity while also showing a strong field enhancement. The difference in absolute value between the RS vectors can be seen as a measure for the chirality density inside the cavity.   The cavity itself is however not chiral as \textit{any} elliptical input polarization is supported. The magnetic field $\mathbf H (\mathbf R)$ is determined via Maxwell's equations from the excitations of the electric dipoles on the metasurface as detailed in App.~\ref{sec:AppendixA}.

The transmission profile around the cavity resonance is illustrated in Fig.~\ref{fig:fig1}(d) for different cavity lengths for a lattice spacing of $a=0.8\lambda$ where the collective dipole shift is close to zero, i.e., $\wt\Omega(0)\approx 0$. If the cavity length $\ell$ exactly matches $n\lambda/2$ ($n\in\mathbb{N}$), the cavity resonance coincides with the resonance of the individual arrays and no transmission is obtained as all the light is reflected. If $\ell$ becomes slightly larger, a narrow transmission window opens up as the mirrors and the cavity now possess different resonance frequencies. Further increasing the cavity length leads to a strongly asymmetric Fano-type profile with a larger linewidth and the cavity resonance drifting towards infinity for $\ell\to(n+\frac{1}{2})\lambda/2$. Once the next multiple of $\lambda/2$ is approached, the cavity linewidth becomes narrow again and the cavity resonance shifts towards $\omega_\nu+\wt\Omega (0)$ as $\tan (k_l\ell)\to 0$. The distance between the zero and the maximum of the transmission can be used as a measure for the cavity linewidth $\kappa =\wt\Gamma (0) |\tan(k_l\ell)|$.  We present a coupled-modes theory for the input-output description of cavities made from quantum metasurface mirrors in App.~\ref{sec:AppendixG}. \\

\begin{figure*}[t]
  \centering
  \includegraphics[width=\textwidth]{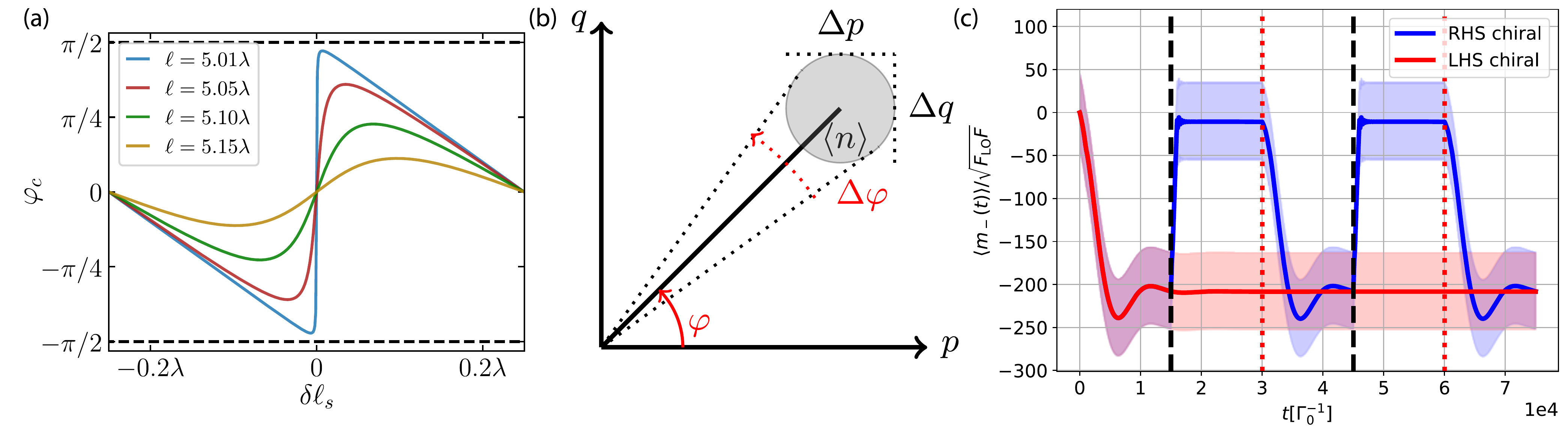}
  \caption{\textit{Chiral sensing.} (a) Cavity phase $\varphi_c=\mathrm{arg}(t_c)$ versus small perturbations of the cavity length $\delta\ell_s$ for different cavity lengths $\ell$ (square lattice, $a=0.8\lambda$). For each $\ell$, the resonance condition described by Eq.~\eqref{eq:resonance} is fulfilled. (b) Illustration of phase uncertainty in phase space for a coherent state. From a geometric point of view, it becomes clear that the phase uncertainty becomes smaller when the amplitude of the field is larger. (c) Homodyne detection signal $\expval{m_- (t)}$ for (ideal) right- and left-handed chiral scatterers (RHS/LHS) passing through an RCP cavity with corresponding shot noise shown in light blue and red, respectively. The dashed black line indicates the entry of a particle into the cavity and the dotted red line indicates its exit. In addition, a rotational average was performed as detailed in App.~\ref{sec:AppendixH}. We have used $\Delta_s=10\Gamma_0$, $\gamma_s=\Gamma_0$ with an average of one photon within the time interval $\Gamma_0^{-1}$, i.e., $F=\Gamma_0$, and an integration time of $T\Gamma_0=2000$. The linewidth for the metasurfaces was assumed to be the same as for the infinite square lattice at $a=0.8\lambda$. The detuning of the incoming laser was chosen to be $\Delta=\Gamma_0/100$.}
  \label{fig:fig2}
\end{figure*}

\noindent \textbf{Chiral sensing} -- We now consider the scenario depicted in Fig.~\ref{fig:fig1}(e) where chiral scatterers with radiative linewidth $\gamma_s$ are sent through the cavity. We assume the resonance of the scatterer $\omega_s$ to be far-detuned from the cavity resonance $\abs{\Delta_s}\gg\abs{\Delta_c}, \gamma_s$ with $\Delta_{s/c}=\omega_l-\omega_{s/c}$. In this case, the effect of a scatterer with the same helicity as the cavity is to increase the path length of light passing through the cavity and thereby effectively shift the cavity length by a small amount $\delta\ell_s=-\arctan(\gamma_s/2\Delta_s)/k_l$ (for derivation see App.~\ref{sec:AppendixF}), such that the total cavity length is now given by $\ell+\delta\ell_s$. On the other hand, a scatterer with the opposite helicity as the cavity mode does not cause a shift (assuming ideal chiral scatterers, in reality both helicities will lead to differential shifts). Due to the quick phase switch around the cavity resonance for cavity lengths close to $n\lambda/2$, a small perturbation of the cavity length can lead to a considerable phase shift of the cavity transmission (see Fig.~\ref{fig:fig2}(a)).  This is the central idea of the sensing scheme discussed in the following. 

One can then proceed to compute the relative phase change in the cavity transmission between lengths $\ell$ and $\ell+\delta\ell_s$ on the cavity resonance (assuming $\ell\approx n\lambda/2$)
\begin{equation}
  \label{eq:ratio}
  \varphi=\arg \frac{t_c(\ell+\delta\ell_s)}{t_c(\ell)}\Bigr|_{\text{res},\ell\approx n\frac{\lambda}{2}}\approx\arctan(\frac{1}{\tan(k_l\delta\ell_s)}),
\end{equation}
which reaches a value of $\pi[\theta(k_l\delta\ell_s)-1/2]$ as $\delta\ell_s\to 0$, implying a phase jump from $-\pi/2$ to $\pi/2$ around the cavity resonance for lengths close to $n\lambda/2$ ($\theta(x)$ is the Heaviside function).  If the cavity length departs from $n\lambda/2$, the cavity linewidth increases and the phase switch gets diminished, as illustrated in Figs.~\ref{fig:fig1}(d) and \ref{fig:fig2}(a).

An experimental setup to measure this phase is homodyne detection as illustrated in Fig.~\ref{fig:fig1}(e) where the phase between a local oscillator (for instance obtained from beam splitting the input field) is compared to the phase of the output field. Suppose we consider a signal beam with which we drive the cavity $\alpha(t)=\sqrt{F}\exp(-\iu\omega_0t+\iu\theta)$ and the local oscillator field $\alpha_L(t)=\sqrt{F_\mathrm{LO}}\exp(-\iu\omega_0t+\iu\theta_\mathrm{LO})$ with intensities (number of photons per unit of time) $F$ and $F_\mathrm{LO}$. Then considering homodyne detection for a Fabry-P\'erot cavity leads to an uncertainty in the measured phase for an integration time $T$ of the measurement and a quantum efficiency $\eta_Q$ which is encoded in the intensity difference $m_-$ with variance $(\Delta m_-)^2_\text{res}= \eta_Q TF_\mathrm{LO}$ and expectation value (assuming the phase variation to happen on a timescale much slower than the optical frequency)
\begin{equation}
  \label{eq:phase_uncertainty}
  \expe{m_-(t,T)} = 2\eta_Q \sqrt{F  F_\mathrm{LO}}\abs{t_c}\int_{t}^{t+T}\diff{t'}\sin \left(\varphi(t') \right).
\end{equation}
Here, we have approximated that $F_\mathrm{LO}\gg F$ and have taken $\theta+\theta_\mathrm{LO}=2\pi n$ which can be obtained by phase matching the local oscillator and the signal beam. Since the cavity is a linear element, the resulting phase uncertainty (see sketch in Fig.~\ref{fig:fig2}(b)) is independent of any cavity properties and only depends on the properties of the state of the incoming beam, which is assumed to be classical for this calculation. This uncertainty could however be improved upon by choosing a phase-squeezed input field instead of a coherent one. 

As an alternative to describing the passage of particles through the cavity with transfer matrix theory, the dipole theory can be extended to include the presence of an additional chiral scatterer which can be represented by coupled electric and magnetic dipoles (see App.~\ref{sec:AppendixH}). These equations of motion are simulated in Fig.~\ref{fig:fig2}(c) for right- and left-handed scatterers (RHS/LHS) entering an RCP cavity, showing a clear distinction in the resulting signal. Here, the classical shot noise of a coherent input field is used to estimate the phase error for a homodyne detection. One can also see several aspects of the cavity physics in this plot. First, the relaxation to the steady state occurs very slowly owing to the very small decay rate of the hybrid cavity. We can also observe that with sufficient detector integration the shot noise can be overcome in order to detect a single scatterer.

Let us finally briefly discuss the applicability of the chiral sensing scheme to the discrimination of molecular enantiomers. Chiral molecules are in general not perfect chiral scatterers. This is manifested in the fact that their circular dichroism $\text{CD}=(A_+-A_-)/(A_++A_-)$, i.e., the difference in absorbance between RCP and LCP light, is not unity but some small finite value. This can have several reasons, but the most physical one is that the magnetic dipole linewidth of an optical transition is usually much weaker than the electric dipole linewidth. This implies that the assumption of coupling to only a single polarization component is unrealistic. Instead, both enantiomers will lead to a small differential shift in optical path length. Aside from the absolute magnitude of this change in path length however, the presented strategy retains generality and the applicabilty is a question of detailed system parameters. More so, we claim that we have mapped the problem of chirality sensing of entantiomers onto a controllable cavity optomechanical setup.\\

\noindent \textbf{Conclusions and Outlook} - 
We have shown that HP mirrors and cavities can be created from stacked quantum metasurfaces with orthogonal dipole orientation. We remark that our proposal could be analogously implemented with arrays of classical dipoles such as plasmonic lattices in which case no external magnetic field control would be needed as the polarization can be controlled by the geometry of the individual plasmonic elements \cite{choudhary2023dark}. We then proposed to use these narrow-linewidth HP cavity modes for the optical sensing of chiral scatterers by discussing how the phase of the output field is modified by an off-resonant scatterer passing through the cavity. We furthermore discussed the phase uncertainty for homodyne detection which can be minimized by tuning the input intensity and the integration time of the detector. We also discussed briefly the applicability to the discrimination of molecular enantiomers. Future endeavors will see the extension of our formalism to additionally describe possible derogating effects such as motion, vacancies and nonlinearities of the quantum emitter array.  \newline\indent  We remark that, in addition to the use proposed in this work, layered metasurfaces can enable a host of other applications. For instance, stacking many of these layers leads to Bragg-mirror physics which could be used to tailor the frequency windows of optical elements based on quantum metasurfaces.  Also, tilting the metasurfaces with respect to each other gives rise to moir\'e superlattices which are known to exhibit exotic optoelectronic phenomena in solid-state platforms \cite{sunku2018photonic}. Preliminary calculations show however that, for normally incident light, the twisting angle between the layers does not matter. Even more general polarization structures of the cavity mode, such as Faraday cavities~\cite{huebener2021strong}, might also be implementable.  \\

\noindent \textbf{Acknowledgments} --  We acknowledge fruitful discussions with L.~Mauro and J.~Fregoni which led to the initial idea for this project. This work was supported by the Max Planck Society and the Deutsche Forschungsgemeinschaft (DFG, German Research Foundation) -- Project-ID 429529648 -- TRR 306 QuCoLiMa
(``Quantum Cooperativity of Light and Matter'').

\FloatBarrier
\bibliographystyle{apsrev4-1-custom}
\bibliography{references}

\onecolumngrid

\appendix
\newpage

\setcounter{secnumdepth}{2}

\section{Dyadic Green's function}\label{sec:AppendixA}

The free-space photonic Green's tensor is given by (evaluated at the resonance of the emitters $\omega_0=ck_0$)
\begin{align}
  \label{definitiongreen_app}
  \mathbf{G}(\mathbf{R})=\left(\mathds{1}+\frac{1}{k_0^2} \nabla\otimes\nabla\right)\frac{\me^{\mi k_0R}}{4\pi R}-\frac{\mathds{1}}{3k_0^2}\delta(\mathbf{R}),
\end{align}
where $\otimes$ denotes the dyadic product, $R=|\mathbf{R}|$ and the last term removes the divergence on the self-interaction terms at $\mathbf{R}=0$. This can be expressed more explicitly as
\begin{align}
  \mathbf{G}(\mathbf{R})=\frac{\me^{\mi k_0 R}}{4\pi k_0^2}\left[\left(\frac{k_0^2}{R}+\frac{\mi k_0}{R^2}-\frac{1}{R^3}\right)\mathds{1}+\left(-\frac{k_0^2}{R}-\frac{3\mi k_0 }{R^2}+\frac{3}{R^3}\right)\frac{\mathbf{R}\otimes\mathbf{R}}{R^2}\right]-\frac{\mathds{1}}{3k_0^2}\delta(\mathbf{R}),
\end{align}
where $\mathds{1}$ is the $3\times 3$ identity matrix  and $\mathbf{R}=(x,y,z)^\top$. For the scattering problem considered here, a Fourier decomposition of the Green's tensor with respect to the in-plane wave vector components $\bq=(q_x,q_y)^\top$ is useful. Decomposing also $\mathbf R=(\br_\parallel,z)^\top$ into lattice plane and out-of-plane coordinate, one can make use of the Weyl expansion \cite{novotny2006principles}
\begin{align}
  \frac{\me^{\mi k_0 R}}{R}=\frac{\mi}{2\pi}\int \dd \bq\, \frac{1}{q_z}\me^{\mi \bq\cdot \br_\parallel}\me^{\mi q_z |z|},
\end{align}
with $q_z=\sqrt{k_0^2-q^2}$. The Green's tensor (including the divergent self-interaction) can now be written as
\begin{align}
  \label{eq:greenfourier}
  \mathbf{G}(\mathbf{R})=\frac{\mi}{8\pi^2 k_0^2}\int \dd\bq \,\frac{k_0^2\mathds{1}-\bar\bq\otimes\bar\bq}{q_z}\me^{\mi \bq\cdot \br_\parallel}\me^{\mi q_z |z|},
\end{align}
where $\bar\bq=(q_x,q_y,\text{sgn}(z)\sqrt{k_0^2-q^2})^\top$. 

From Maxwell's equations for monochromatic fields, one can find the magnetic Green's tensor (including the self-interaction) as
\begin{equation}
  \label{eq:magnetic_GF}
  \mathbf G_M(\mathbf R)=\frac{\iu}{\omega_0}\curl\mathbf G(\mathbf R)=\frac{\iu}{4\pi\omega_0}\curl\left(\mathds 1 \frac{\me^{\iu k_0 R}}{R}\right)=\frac{(k_0 R+\iu)\me^{\iu k_0 R}}{4\pi\omega_0 R^3}
  \begin{pmatrix}
    0 & -z & y\\
    z  & 0 & -x\\
    -y  & x & 0
  \end{pmatrix}.
\end{equation}
which can be used to determine the magnetic field emitted from an electric dipole. Here, the curl of a matrix $\mathbf A$ is defined (using Einstein sum convention) as $(\curl \mathbf A)_{il}=\varepsilon_{ijk}\partial_j A_{kl}$ with $\varepsilon_{ijk}$ the Levi-Civita symbol.

\section{Fourier space treatment of bilayer system: Intra- and interlayer interactions}
\label{sec:AppendixB}

As mentioned in the main text, the HP cavity setup can be decomposed into two cavities consisting of two metasurfaces each. We will thus restrict our discussion here to bilayers. The bilayer system forming the cavity can be treated analytically by a Fourier transform of the dipole operators on both left ($L$) and right ($R$) sublattice. We define the Fourier transformation for the Pauli operators (considering a single polarization degree of freedom) as
\begin{subequations}
  \begin{align}
    \sigma_\bq^{L/R}=\sum_{j\in L/R}\sigma_j \me^{-\mi \bq \cdot\br_j}, \qquad \sigma_j^{L/R}=\frac{1}{\calN}\sum_\bq \sigma_\bq^{L/R} \me^{\mi\bq\cdot \br_j^{L/R}}.
  \end{align}
\end{subequations}
The effective non-Hermitian Hamiltonian describing the free propagation of a single excitation on the two independent arrays may then be expressed in Fourier space as
\begin{align}
  \label{eq:heff}
  \widetilde \calH\ts{eff}=\sum_{\bq}\left(-\Delta+\widetilde\Omega(\bq)-\mi\frac{\widetilde\Gamma (\bq)}{2}\right)\left(\sigma_\bq^{R, \dagger}\sigma_\bq^{R}+\sigma_\bq^{L, \dagger}\sigma_\bq^{L}\right),
\end{align}
where $\widetilde\Omega(\bq)$, $\widetilde\Gamma (\bq)$ denote the in-plane frequency shifts and decay rates, respectively (definition see below in Eq.~\eqref{eq:inplane}). Importantly however, in addition to Eq.~\eqref{eq:heff}, one is left with a term describing the interlayer interaction between the arrays which is discussed in more detail below. The laser drive under general oblique incidence with respect to the array plane is included as
\begin{align}
  \calH_l=\sum_{j\in L}\left(\eta^L\me^{\mi\bk_\parallel\cdot\br_j}\sigma_j^\dagger+\mathrm{h.c.}\right)+\sum_{j\in R}\left(\eta^R\me^{\mi\bk_\parallel\cdot\br_j}\sigma_j^\dagger+\mathrm{h.c.}\right),
\end{align}
where $\bk_\parallel$ is the wave vector of the laser parallel to the array plane, $k_z=\sqrt{k_l^2-k_\parallel^2}$ and the Rabi drive for the two arrays differs by a phase factor due to the free propagation of the incoming field along the $z$-direction $\eta^R=\eta^L\me^{\mi k_z\ell}$. 
The linearized equations of motion for the expectation values of the dipole amplitudes $\beta_\bq^{L/R}=\expval{\sigma_\bq^{L/R}}$ can be expressed  in the weak excitation limit $\expval{\sigma_j^z}\approx -1$  in Fourier space as
\begin{subequations}
\label{eq:sublattice_eom}
  \begin{align}
    \dv{\beta^L_\bq}{t}&=\iu(\Delta-\calM_\bq)\beta^L_\bq-\iu\calM^{LR}_\bq\beta^R_\bq-\iu\calN\delta_{\bq,\bk_\parallel }\eta^L,\\
    \dv{\beta^R_\bq}{t}&=\iu(\Delta-\calM_\bq)\beta^R_\bq-\iu\calM^{LR}_\bq\beta^L_\bq-\iu\calN\delta_{\bq,\bk_\parallel }\eta^R,
  \end{align}
\end{subequations}
from which we can see that an incident laser only probes the surface modes corresponding to $\bq=\bk_\parallel$. The term $\calM_\bq=\wt\Omega (\bq)-\mi\wt\Gamma (\bq)/2$ describing the collective frequency shifts and decay rates of an individual array and is given by a sum over all in-plane interactions (for an arbitrary index $0$ on one of the two arrays)
\begin{align}
\label{eq:inplane}
  \calM_\bq =\sum_{j\in \text{in-plane}} \left(\Omega_{0j}-\mi\frac{\Gamma_{0j}}{2}\right)\me^{-\mi\bq\cdot\br_j}= \widetilde\Omega(\bq)-\mi\frac{\widetilde\Gamma (\bq)}{2}=-\frac{3}{2}\Gamma_0\lambda_0\wt G(\bq;0),
\end{align}
which can be expressed in terms of the Fourier transform of the Green's tensor (single polarization component) with respect to its first two arguments $\wt G (\bq;z)=\sum_{j}\me^{-\mi\bq\cdot\br_j}G(\mathbf R)$. The term $\calM^{LR}_\bq$, which describes the interaction between the two arrays, involves a summation over all coherent and dissipative out-of-plane interactions with the emitters of the opposing lattice
\begin{align}
  \calM_\bq^{LR}=\sum_{j\in \text{out-of-plane}} \left(\Omega_{0j}-\mi\frac{\Gamma_{0j}}{2}\right)\me^{-\mi\bq\cdot\br_j}=-\frac{3}{2}\Gamma_0\lambda_0\wt G(\bq;\ell),
\end{align}
which simply corresponds to evaluating the 2D lattice transform of the Green's tensor at $z=\ell$.
In steady state, the equations of motion for the coherences \eqref{eq:sublattice_eom} become
\begin{subequations}
  \begin{align}
    \label{eq:plugged_into}
    \left[\Delta-\calM_\bq-\frac{\left(\calM_\bq^{LR}\right)^2}{\Delta-\calM_\bq}\right]\beta^L_\bq&=\calN\delta_{\bq,\bk_\parallel }\left(\eta^L+\frac{\calM_\bq^{LR}\eta_\bq^R}{\Delta-\calM_\bq}\right),\\
    \left[\Delta-\calM_\bq-\frac{\left(\calM_\bq^{LR}\right)^2}{\Delta-\calM_\bq}\right]\beta^R_\bq&=\calN\delta_{\bq,\bk_\parallel }\left(\eta^R+\frac{\calM_\bq^{LR}\eta_\bq^L}{\Delta-\calM_\bq}\right).
  \end{align}
\end{subequations}
By making use of Poisson's summation formula, the Fourier transform of the Green's tensor can be turned into a sum over all vectors $\mathbf g$ of the reciprocal lattice $\Lambda^*$ and can be approximated in the far field $z\gg\lambda$ as
\begin{align}\label{eq:reciprocal_lattice_green}
  \wt G (\bq;z)=\frac{\mi}{2\calA}\sum_{\mathbf g\in\Lambda^*}\frac{\me^{\mi\sqrt{k_0^2-(\bq+\mathbf g)^2}|z|}}{\sqrt{k_0^2-(\bq+\mathbf g)^2}}\approx \frac{\mi}{2\calA}\frac{\me^{\mi q_z |z|}}{q_z},
\end{align}
where we have only kept the $\mathbf g=0$ contribution and took all other contributions to be evanescent as it is the case for subwavelength lattices.
The interaction between the arrays is then simply described by the plane wave term
\begin{align}
  \calM_\bq^{LR}= -\mi\frac{\wt \Gamma (\bq)}{2}\me^{\mi q_z\ell},
\end{align}
with the amplitude of the interaction is governed by the effective in-plane decay rate $\widetilde\Gamma (\bq)=3\Gamma_0 /(4\pi )(\lambda_0^2/\calA) (k_0/q_z)$. This leads to the following expressions for the dipole amplitudes on left and right layer:
\begin{align}
  \label{eq:plugged_into_cavity}
  \beta^L_\bq=\calN\delta_{\bq,\bk_\parallel }\eta^L\left[\frac{\Delta-\calM_\bq-\mi\frac{\wt\Gamma (\bq)}{2}\me^{2\mi q_z\ell}}{(\Delta-\calM_\bq)^2+\frac{\wt\Gamma (\bq)^2}{4}\me^{2\mi q_z\ell}}\right],\qquad \beta^R_\bq=\calN\delta_{\bq,\bk_\parallel }\eta^L\left[\frac{\left(\Delta-\calM_\bq-\mi\frac{\wt\Gamma (\bq)}{2}\right)\me^{\mi q_z\ell}}{(\Delta-\calM_\bq)^2+\frac{\wt\Gamma (\bq)^2}{4}\me^{2\mi q_z\ell}}\right].
\end{align}

\section{Transmission, intracavity field, and resonance condition}\label{sec:AppendixC}

The total electric field can be computed as the sum of the incident field and the dipole-scattered field $E^{(+)}(\mathbf R)=E\ts{in}^{(+)}\me^{\mi k z}+E\ts{dip}^{(+)}(\mathbf R)$. The dipole field is given by the sum over all individually emitted fields and can be expressed as a sum over all wave vectors as (only considering the field along the $z$-direction)
\begin{equation}\label{eq:A_scattered_field}
  E^{(+)}_{\text{dip}}(z)=-\frac{3\pi\Gamma_0}{k_0d }\sum_j G(z\hat e_z-\br_j)\beta_j=-\frac{3\pi\Gamma_0}{k_0d \calN}\sum_\bq\left[\widetilde{ G}^L(\mathbf q;z)\beta_{\mathbf q}^L+\widetilde{ G}^R(\mathbf q;z)\beta_{\mathbf q}^R\right],
\end{equation}
where $\widetilde{ G}^L$, $\widetilde{ G}^R$ denote the Fourier transforms of the Green's tensor with respect to the left and right sublattice.
This leads to the following expressions for the transmitted and intracavity field (neglecting all contributions from evanescent terms)
\begin{subequations}
  \begin{align}
    \label{eq:lattice_ft_zero}
    E^{(+)}(z >\ell)&=\left[E_{\text{in}}^{(+)}\me^{\iu k_z z}-\iu\frac{\widetilde \Gamma(\bk_\parallel)}{2d}\left(\beta^L_{\bk_\parallel}+\me^{-\iu q_z\ell}\beta^R_{\bk_\parallel}\right)\me^{\iu q_z z}\right]\approx t_c E_{\text{in}}^{(+)}\me^{\iu k_z z},\\
    E^{(+)}(0<z<\ell)&=\left[E_{\text{in}}^{(+)}\me^{\iu k_z z}-\iu\frac{\widetilde \Gamma({\bk_\parallel})}{2d}\left(\beta^L_{\bk_\parallel}\me^{\iu q_z z}+\me^{\iu q_z(\ell-z)}\beta^R_{\bk_\parallel}\right)\right],
  \end{align}
\end{subequations}
\begin{figure}[t]
  \centering
  \includegraphics[width=0.5\columnwidth]{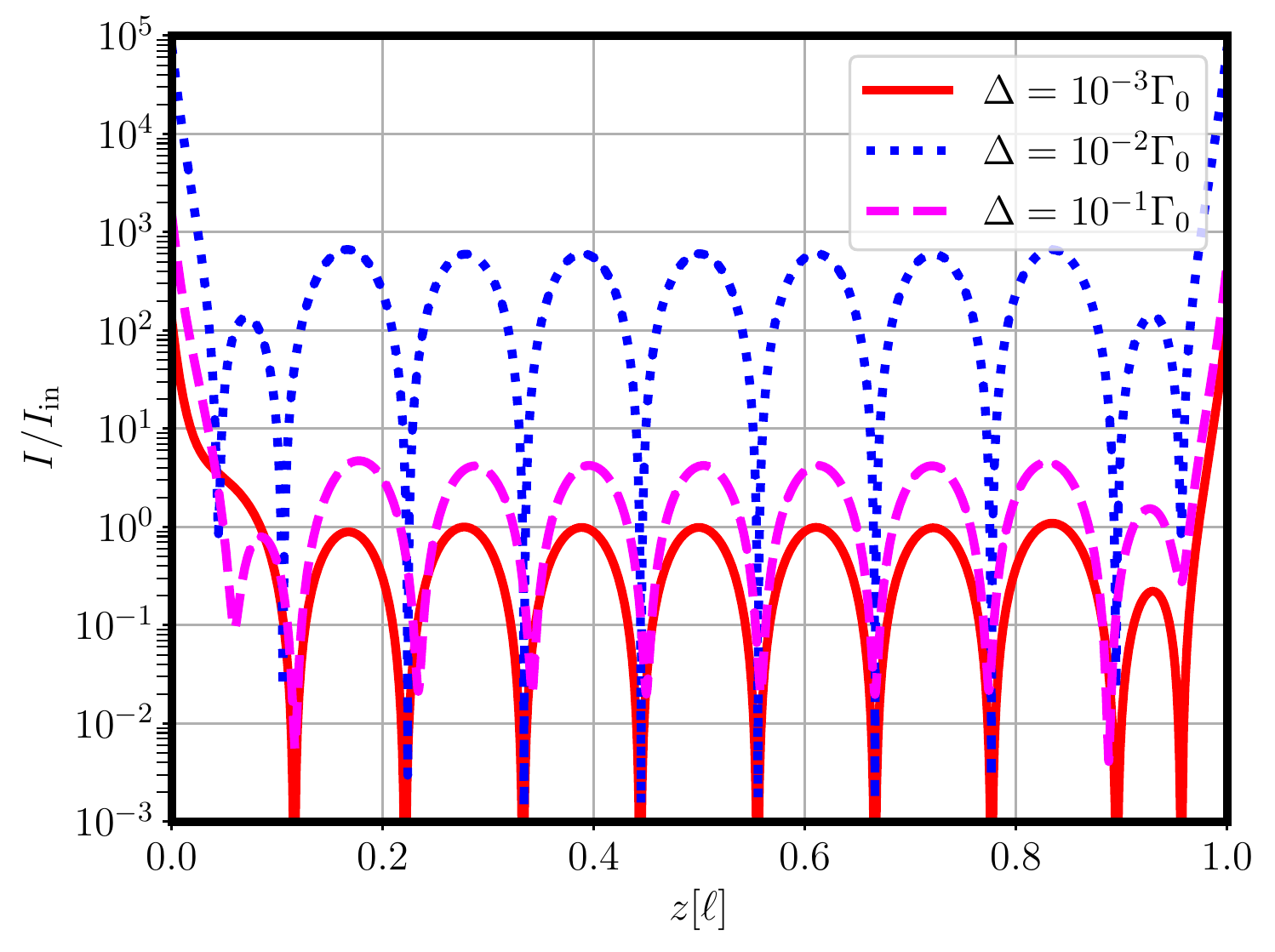}
  \caption{Plots of intracavity field intensity for a bilayer system where the resonance condition for $\Delta$ is always satisfied. The $z$-coordinate is normalized to the cavity length so that the different cases can be compared. The plots are generated by considering the infinite square lattice with $a=0.8\lambda$ in Fourier space. Eq.~\eqref{eq:reciprocal_lattice_green} was evaluated until the reciprocal lattice sum converged so that evanescent contributions to the interlayer interaction and cavity field are also included.}
  \label{fig:cavity_field}
\end{figure}
where the cavity transmission coefficient reads
\begin{align}
  \label{eq:transcoeff}
  t_c=\frac{\left[\Delta-\calM_{\bk_\parallel}-\mi\frac{\wt\Gamma ({\bk_\parallel})}{2}\right]^2}{(\Delta-\calM_{\bk_\parallel})^2+\frac{\wt\Gamma ({\bk_\parallel})^2}{4}\me^{2\mi k_z\ell}},
\end{align}
and we have assumed $q_z\approx k_z$. The condition that the cavity transmission should equal unity at the cavity resonance $|t_c|^2=1$ yields a quadratic equation for the collective resonance $\Delta-\wt\Omega(\bk_\parallel)$ with a single solution given by
\begin{align}
  \Delta-\wt\Omega(\bk_\parallel)=-\frac{\wt\Gamma({\bk_\parallel})}{2}\frac{\sin(2k_z \ell)}{1+\cos(2 k_z\ell)}=-\frac{\wt\Gamma({\bk_\parallel})}{2}\tan(k_z\ell),
\end{align}
implying that the resonance frequency of the cavity is given by $\omega_c=\omega_0+\wt\Omega (\bk_\parallel)-\wt\Gamma({\bk_\parallel})\tan(k_z\ell)/2$. The results presented in the manuscript correspond to $\bk_\parallel = 0$, i.e., perpendicular illumination of the cavity. In Fig.~\ref{fig:cavity_field} we plot the intensity of the intracavity field at different resonances including the contribution of the evanescent terms, leading to a field distribution which is maximal at the mirror surfaces and exhibiting a sine-like behavior in between.  

The results can be easily generalized to several polarization components (e.g., in the Cartesian polarization basis)  by considering the vector of dipole amplitudes $\boldsymbol \beta_\bq=(\beta_{\bq,x},\beta_{\bq,y},\beta_{\bq,z})^\top$ with $\beta_{\bq,\nu}=\expval{\sigma_{\bq,\nu}}$ and by taking into account the full Green's tensor Eq.~\eqref{eq:greenfourier}. 

\section{Comparison of Fourier space solution with finite-size real space simulation}\label{sec:AppendixD}

\begin{figure}[h]
  \centering
  \includegraphics[width=0.99\columnwidth]{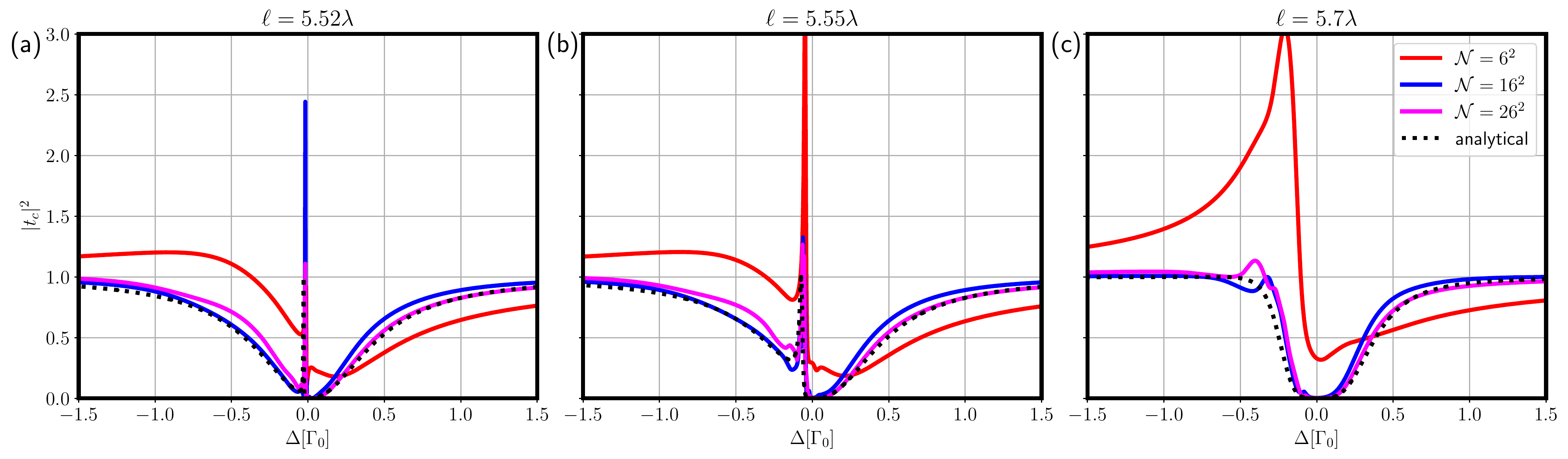}
  \caption{Comparison between analytical expression of transfer matrix/Fourier space transmission lineshapes (Eq.~\eqref{eq:transcoeff}, dotted black curve) with real space simulation (solid curves) for square lattices with $a=0.8\lambda$ for different cavity lengths (a) $\ell=5.52\lambda$, (b) $\ell=5.55\lambda$, and (c) $\ell=5.7\lambda$. The number of emitters per metasurface is given by $\calN$. For the numerical simulations, a small array curvature with radius $R\approx 3.5\cdot 10^8\lambda$ was implemented since the finite size of the array would otherwise lead to instabilities due to transverse waveguiding modes \cite{guimond2019subradiant}. Simulations for other curvature radii show similar results.}
  \label{fig:finite_size}
\end{figure}

The analytical results derived above are valid for infinite lattices. In Figs.~\ref{fig:finite_size}(a)-(c), we compare the result of Eq.~\eqref{eq:lattice_ft_zero} with the transmission profile obtained by a real space simulation of the coupled-dipole equations for a finite number of emitters on the layers, for different cavity lengths. For a finite lattice, also other modes than $\bq=0$ are excited by the incoming laser. The scattering contributions from these modes lead to a transmission larger than unity and vanish for larger array sizes, converging towards the analytical solution. While the results presented below correspond to plane wave illumination, for a Gaussian beam with beam width covering only a few lattice sites, the infinite lattice case is approached more quickly as boundary effects are avoided.

\section{Transfer matrix for HP mirror}\label{sec:AppendixE}

In this section, we show that the proposed setup indeed preserves the helicity by making use of the transfer matrix method. We start by constructing the transfer matrix for a single HP mirror consisting of two metasurface layers with dipoles oriented along $x$ and $y$. From this, the cavity properties can then be determined in a straightforward manner. Consider the setup of mirrors with corresponding transfer matrices of
  \begin{equation}
    \label{eq:transfer_mirror}
    \mathbf{ T}_m=\begin{pmatrix}
      1+\iu\zeta_m & \iu\zeta_m\\
      -\iu\zeta_m & 1-\iu\zeta_m
    \end{pmatrix},
  \end{equation}
  where $\zeta_m$ is the polarizability of the metasurface. The first mirror only affects the $x$ polarization and the second only the $y$ polarization component. Between them the light propagates freely according to
  \begin{equation}
    \label{eq:free_space_propagation}
    \mathbf{ T}_f=\begin{pmatrix}
      \me^{\iu k\ell_m} & 0\\
    0 & \me^{-\iu k\ell_m}
    \end{pmatrix},
  \end{equation}
  for both polarization components. In the basis $(E^\leftarrow_x,E^\rightarrow_x,E^\leftarrow_y,E_y^\rightarrow)^\top$, where the subscript dennotes the polarization and the superscript the propagation direction of the electric field, the total transfer matrix can be expressed in block notation as
  \begin{equation}
    \label{eq:transfer_total}
    \mathbf { T}\ts{HP}=\begin{pmatrix}
      \mathbf{ T}_m & 0\\
      0 & \mathds 1
    \end{pmatrix}
    \begin{pmatrix}
      \mathbf{ T}_f & 0\\
      0 & \mathbf{ T}_f
    \end{pmatrix}
    \begin{pmatrix}
      \mathds 1 & 0\\
      0 & \mathbf{ T}_m
    \end{pmatrix}.
  \end{equation}
 To see now that this matrix represents a HP mirror for $\ell_m=\lambda/4$, we must perform a basis transformation which rotates $(E^\leftarrow_x,E^\rightarrow_x,E^\leftarrow_y,E_y^\rightarrow)^\top$ into $(E^\leftarrow_+,E^\rightarrow_+,E^\leftarrow_-,E_-^\rightarrow)^\top$. This is given by

  \begin{equation}
    \label{eq:basis_transform}
    \mathbf U =\frac{1}{\sqrt{2}}
    \begin{pmatrix}
      1 & 0 & 1 & 0 \\
      0 & 1 & 0 & 1 \\
      -\iu & 0 & \iu & 0 \\
      0 & \iu & 0 & -\iu
    \end{pmatrix}.
  \end{equation}
  Applying this basis transformation to the transfer matrix yields
  \begin{equation}
    \label{eq:basis_transformed_matrix}
    \mathbf U^\dagger\mathbf { T}\ts{HP}\mathbf U=
    \begin{pmatrix}
      1+\iu \zeta_m  & \iu \zeta_m  & 0 & 0 \\
      -\iu \zeta_m  & 1-\iu \zeta_m  & 0 & 0 \\
      0 & 0 & 1+\iu \zeta_m  & \iu \zeta_m  \\
      0 & 0 & -\iu \zeta_m  & 1-\iu \zeta_m
    \end{pmatrix}
    \begin{pmatrix}
      \iu & 0 & 0 & 0\\
      0 & -\iu & 0 & 0\\
      0 & 0 & \iu & 0\\
      0 & 0 & 0 & -\iu
    \end{pmatrix},
  \end{equation}
  which is indeed block-diagonal in the circular basis with identical subblocks for the two polarization components. The diagonal matrix on the right-hand side accounts for the free propagation. In total, it can thus be formulated that the resulting optical element describes a ``\textit{helicity-preserving mirror plus free propagation of $\lambda/4$}''.
  
\section{Cavity phase shift due to presence of chiral scatterer}\label{sec:AppendixF}

For further discussions, we focus on one of the equivalent subblocks in Eq.~\eqref{eq:basis_transformed_matrix}. A chiral molecule can be interpreted as a scatterer that induces different path lengths for the two polarization components $\delta l_s^{\pm}$.

The transfer matrix for a single scatterer expresses as
\begin{align}
  \mathbf{T}_s =
  \begin{pmatrix}
    1+\mi\zeta_s & \mi\zeta_s\\
    -\mi\zeta_s & 1-\mi\zeta_s
  \end{pmatrix},
\end{align}
where the polarizability of a single scatterer with linewidth $\gamma_s$ and detuning from the laser frequency $\Delta_s=\omega_l-\omega_s$ is given by $\zeta_s=-(\gamma_s/2)/(\mi\gamma_s/2+\Delta_s)$. In the limit $\abs{\Delta_s}\gg\gamma_s$, this can be understood as an effective path length added to the free space propagation inside the cavity due to the approximation
\begin{align}
  \mathbf{T}_s \approx
  \begin{pmatrix}
    \me^{\mi k_l\delta\ell_s} & 0\\
    0  & \me^{-\mi k_l \delta\ell_s}
  \end{pmatrix},
\end{align}
with the effective length change of the cavity caused by the scatterer
\begin{align}
  \delta\ell_s=-\frac{\arctan(\gamma_s/2\Delta_s)}{k_l}.
\end{align}
The cavity transmission coefficient is obtained from the last entry of the total transfer matrix $t_c=1/T_{22}$ and reads
\begin{align}
  t_c (\ell+\delta\ell_s)=\frac{\left(\Delta-\wt\Omega (0)\right)^2}{\left(\Delta-\wt\Omega (0)+\mi\frac{\gaz}{2}\right)^2+\frac{\gaz^2}{4}\me^{2\mi k_l (\ell+\delta\ell_s)}},
\end{align}
and the relative phase shift due to the presence of the scatterer in the cavity can be expressed as (on the cavity resonance $\Delta-\wt\Omega (0)=-\frac{\wt\Gamma (0)}{2}\tan (k_l\ell)$)
\begin{align}
  \varphi&=\arg\left[\frac{t_c(\ell+\delta\ell_s)}{t_c(\ell)}\right]_{\text{res}}=-\arccot(\cot(2 k_l\delta\ell_s)-\csc(2 k_l\delta\ell_s)\sec(2 k_l\ell)^2)\\\nonumber
         &\approx \arctan(\cot(2k_l\delta \ell_s))-\frac{k_l^2\bar\ell^2}{2}\cot(k_l\bar\ell),
\end{align}
where the last approximation holds for $\ell\approx n\lambda/2$ and $\bar\ell=\ell-\lfloor 2\ell\rfloor/2$ is the relative distance from this half-integer factor of the wavelength.

Enantiomers can now be differentiated by illuminating the cavity with circularly polarized light. Within the polarization subblocks, the two enantiomers will lead to different effective path lengths $\delta\ell_s^\pm$ and hence to a different signal in the homodyne measurement.

\section{Coupled-modes theory}\label{sec:AppendixG}

\begin{table}[h!]
  \centering
  \begin{tabular}{||l|l|l||}\toprule
    $c_\lambda(\omega,t)$ & bath operator of mode $\lambda$ &\\
     $b_\lambda$ & system oscillator mode $\lambda$ &\\
    $\omega_\lambda$ & frequency of oscillator mode $\lambda$ & \\
    $b_{\text{in},\lambda}$ & input operator of port $\lambda$ & $b_{\text{in},\lambda}(t)=\lim_{t_1\to\infty}\frac{1}{2\pi}\int\diff{\omega}\,\me^{-\iu\omega(t-t_1)}c_\lambda(\omega,t_1)$\\
    $b_{\text{out},\lambda}$ & output operator of port $\lambda$ & $b_{\text{out},\lambda}(t)=\lim_{t_0\to-\infty}\frac{1}{2\pi}\int\diff{\omega}\,\me^{-\iu\omega(t-t_0)}c_\lambda(\omega,t_0)$\\
    $W_{\lambda,\lambda'}$ & coupling between oscillator mode $\lambda$ and input operator $\lambda'$ &\\
    $t_c^\text{TM}$ & transmission function from transfer matrix theory & \\
    $t_c^\text{CM}$ & transmission function from coupled-modes theory & \\
    $\kappa$ & cavity mode linewidth &\\
    $\omega_c$ & cavity mode frequency &\\
    \bottomrule
  \end{tabular}
  \caption{Definitions for section S7.}
  \label{tab:defs7}
\end{table}

The results discussed so far describe the classical light propagation through a hybrid cavity formed by quantum metasurfaces. For many applications (such as sensing discussed below), one is however interested in a quantum input-output description which would also allow to describe non-classical effects such as e.g., the photon statistics of the transmitted field. 

\begin{figure}[b]
  \centering
  \includegraphics[width=\columnwidth]{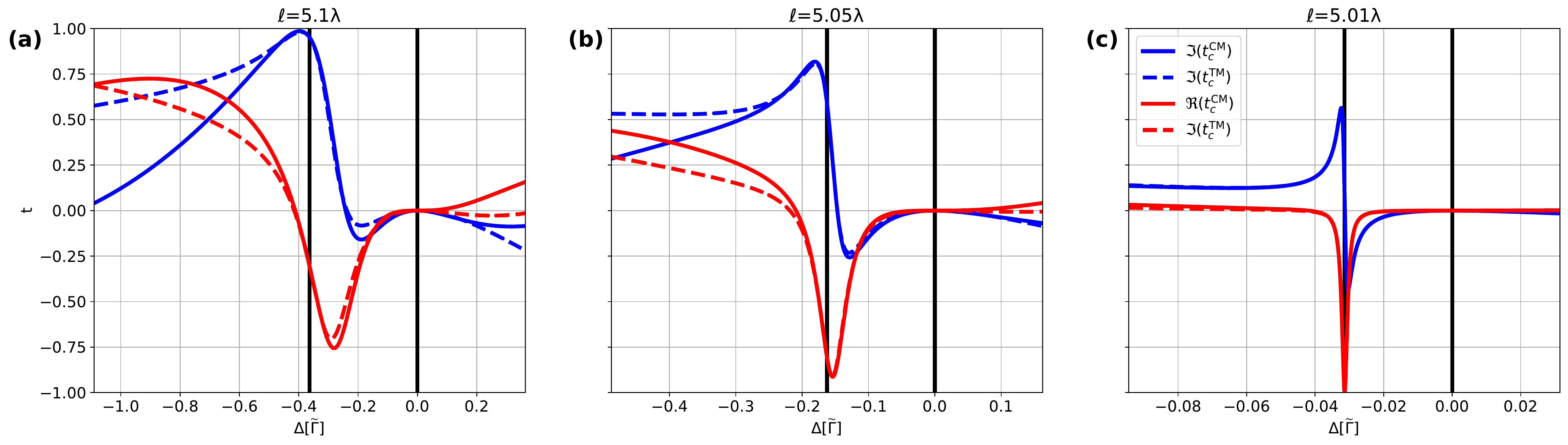}
  \caption{Comparison between cavity transmission function $t_c^\text{TM}$ obtained from transfer matrix theory/coupled-dipole formalism (dashed curves) and the cavity transmission function $t_c^\text{CM}$ of the coupled-modes theory (solid curves). Here we chose $\widetilde\Omega(0)=0$ as the origin. The real part of the transmission functions is shown in red and the imaginary part in blue. From left to right, we show the comparison for decreasing cavity lengths (a) $\ell=5.1\lambda$, (b) $\ell=5.05\lambda$,  and (c) $\ell=5.01\lambda$, which corresponds to increasing reflectivity of the metasurface at the resonance. The transmission is shown within the frequency window $[\omega_c-\kappa,\omega_c+\kappa]$. The vertical line at $\Delta=0$ indicates the metasurface resonance and the other vertical line indicates the cavity resonance.}
  \label{fig:coupled_mode_plots}
\end{figure}

To this end, we assume a phenomenological coupled-modes model (as introduced in Ref.~\cite{cernotik2019cavity}) where the cavity mode and surface modes of the metasurface are modeled by harmonic oscillators coupled to common vacuum modes. We follow the notation in Ref.~\cite{viviescas_field_2003} which assumes equations of motion $\dot b_\lambda=-\iu\omega_\lambda b_\lambda-\iu\sum_{\lambda'}W_{\lambda,\lambda'}b_{\text{in},\lambda}$ for the oscillator modes $b_\lambda$ and input noise operators $b_{\text{in},\lambda}$ (see table of definitions \ref{tab:defs7} above).
  Then, the input-output relations may be formulated in compact matrix-vector form in frequency domain  as
\begin{equation}
  \label{eq:equations_for_coupled_modes_markovian}
    \mathbf b_{\text{out}}(\omega)=\mathbf S(\omega)\mathbf b_{\text{in}}(\omega),
\end{equation}
where the scattering matrix assumes the following form in the Markovian limit
\begin{align}
\mathbf S(\omega)=\mathds{1}-2\pi\iu\mathbf{W}^\dagger\mathbf D^{-1}(\omega)\mathbf{W},
\end{align}
with
\begin{align}
[\mathbf D (\omega)]_{\lambda\lambda'} =(\omega-\omega_\lambda)\delta_{\lambda\lambda'}+\iu\pi[\mathbf{W}\mathbf{W}^\dagger]_{\lambda\lambda'}.
\end{align}
The transmission can then be found as matrix elements of the scattering matrix $t_c^\text{CM}=S_{12}$.

Now, we make a physical assumption for the form of this coupling matrix, see for instance \cite{cernotik2019cavity}
\begin{equation}
  \label{eq:W_form}
  \mathbf{W}=
  \begin{pmatrix}
    \sqrt{\frac{\kappa}{4\pi}} & \sqrt{\frac{\kappa}{4\pi}}\\
    \sqrt{\frac{\widetilde\Gamma(0)}{2\pi}} & 0\\
    0 & \sqrt{\frac{\widetilde\Gamma(0)}{2\pi}}
  \end{pmatrix}.
\end{equation}
By this choice we identify $b_1$ as the cavity mode and $b_2$ and $b_3$ as the surface modes of the metasurface. Thus, we also choose $\omega_2=\omega_3=\omega_0+\widetilde\Omega (0)$. This satisfies the condition that the cavity transmission ought to be zero at the metasurface resonance $t_c^\text{CM}(\omega_0+\widetilde\Omega (0))=0$. The other condition that needs to be satisfied is that the cavity transmission becomes unity at the cavity resonance $ | t_c^\text{CM}(\omega_c) | ^2=1$. It turns out that any choice $\omega_1=\omega_c+\delta$ and $\kappa=\left(\frac{\widetilde\Gamma(0)}{\omega_c}+4\frac{\omega_c}{\widetilde\Gamma(0)}\right)\delta$ will satisfy this condition. The natural choice $\delta=0$ will however not lead to a physical solution. Checking against the transfer matrix theory, we find that the choice $\delta=\omega_c$ gives the correct linewidth.

The comparison between the transfer matrix results and the result for the coupled-modes theory is shown in Fig.~\ref{fig:coupled_mode_plots} for different cavity lengths $\ell$. We can see that for $\ell \approx n\lambda/2$, the coupled-modes theory perfectly fits the transmission around the resonance. For larger deviations from this condition, the transmission functions show differences. This can be interpreted as the contributions of the transmission that occur via the free space modes instead of via the cavity-confined mode for a bad cavity (the transmission of a metasurface becomes unity for large detuning).

\section{Chiral response of far-detuned molecular scatterer in HP metasurface cavity}\label{sec:AppendixH}

\newcommand{\hefac}{\frac{k_0^3}{3\pi\sqrt{\epsilon_0\mu_0}}}
\newcommand{\efac}{\frac{k_0^3}{3\pi\epsilon_0}}
\newcommand{\hfac}{\frac{k_0^3}{3\pi\mu_0}}

\begin{table}[h!]
  \centering
  \begin{tabular}{||l|l|l||}\toprule
    $\omega_s$ & transition frequency of molecule &\\
    $\mathbf{d}_s$ & electric dipole moment of molecule &\\
    $\bm{\mu}_s$ & magnetic dipole moment of molecule &\\
    $\ell_s$ & position of molecule along the $z$ axis &\\
    $\beta_s$ & coherence of molecular transition & $\beta_s=\expval{\ket{e_s}\bra{g_s}}$\\
    $\mathbf d_i$ & dipole moment of $i$th metasurface in stack & \\
    $\beta_i$ & coherence of $i$th metasurface in stack &\\
    $\ell_i$ & position of $i$th metasurface in stack &\\
    $\Delta_s$ & detuning of molecule & $\omega_l-\omega_s$\\
    $\gamma_s$ & total molecule linewidth & $\gamma_s=\gamma^m_s+\gamma^e_s$\\
    $\gamma^m_s$ & molecule magnetic linewidth & $\gamma^m_s=\hfac \mu_s^2$\\
    $\gamma^e_s$ & molecule electric linewidth & $\gamma^e_s=\efac d_s^2$\\
    $\hat e_{\pm}$ & circular basis vectors & $\hat e_\pm=(\hat e_x\pm\iu\hat e_y)/\sqrt{2}$\\
    $\xi$ & polarization of incoming light (RCP/LCP) &$\xi=\pm$\\
    $\expval{\cdot}_{\text{rot}}$ & rotational average & see Ref.~\cite{craig1998molecular}\\
    $\ell_m$ & HP mirror metasurface spacing &\\
    $\beta^x_L$ & coherence of left metasurface with dipole moment in $x$-direction & \\
    $\beta^y_L$ & coherence of left metasurface with dipole moment in $y$-direction & \\
    $\beta^x_R$ & coherence of right metasurface with dipole moment in $x$-direction & \\
    $\beta^y_R$ & coherence of right metasurface with dipole moment in $y$-direction & \\
    $\beta^\xi_L$ & circular superposition of left metasurfaces & $\beta^\xi_L=\beta^x_L+\xi \beta^y_L$\\
    $\beta^\xi_R$ & circular superposition of right metasurfaces & $\beta^\xi_R=\beta^x_R+\xi \beta^y_R$\\
    $\hat{v}$ & a hat indicates that the vector is normalized (unit vector) & $\hat v=\mathbf v/\norm{\mathbf v}$\\
    $\mathcal M_{\mathbf q}$ & interaction between $\mathbf q$-mode of a metasurface and magnetic moment of molecule in far field & \\
    $\mathbf E\ts{in}$ & incoming electric field &\\
    $\mathbf B\ts{in}$ & incoming magnetic field &\\
    $d$ & magnitude of dipole moments of HP mirror metasurfaces & $d=\norm{\mathbf d_i}$\\
    $\widetilde \Gamma$ & linewidth of metasurfaces & $\widetilde \Gamma=\widetilde\Gamma(0)=\efac d^2$\\
    \bottomrule
  \end{tabular}
  \caption{Definitions for section S8.}
  \label{tab:defs8}
\end{table}

Consider now that we place a chiral molecular scatterer with electric moment $\mathbf{d}_s$ and magnetic moment $\bm \mu_s$ inside the cavity. We now assume that both these moments correspond to the same transition, implying cross-terms in the polarizability for the electric and magnetic components, i.e., an electric field will also excite the magnetic dipole moment.

In order to continue, we first need to consider the magnetic far field due to an array of electric dipoles due to the cross term in the Green's function given in Eq.~\eqref{eq:magnetic_GF} (the Green's function for the electric field generated by a magnetic dipole is the same but multiplied by minus one). The procedure to calculate the far field in the subwavelength regime is essentially the same as for the purely electric Green's function. To see this, consider that both Green's functions can be written as a differential operator acting on the scalar Green's function. As such,
\begin{equation}
  \label{eq:Greens_function_magnetic}
  \mathcal M_{\mathbf q} \approx -\iu\frac{k_0^3}{3\pi\sqrt{\epsilon_0\mu_0}}\frac{\bm \mu_s^*\cdot (\hat e_z\times\mathbf d)}{2}\mathrm{sgn}(z)\me^{\iu q_z \abs{z}}
\end{equation}
gives the interaction due to light propagating from the metasurface to the molecule located at position $z$. Here, $\mathbf d$ indicates the dipole moment of the metasurface (also see table of definitions \ref{tab:defs8}). Now, the interaction between $N$ metasurface layers with dipole moments $\mathbf d_i$ and a single molecule can be written as  (denoting for simplicity $  \wt \Gamma (0)\equiv \wt \Gamma$ and omitting the coupling of the molecule to other surface modes of the array with $\mathbf q\neq 0$) 
\begin{subequations}
\begin{align}\label{eq:metasurface_molecule}
  \dv{\beta_i}{t}&=(\iu\Delta-\widetilde\Gamma/2)\beta_i-\efac\sum_{j\neq i}\mathbf d_i^*\cdot\mathbf d_j\me^{\iu k_z\abs{\ell_i-\ell_j}}\beta_j-\iu\mathbf d_i^*\cdot \mathbf E\ts{in}\me^{\iu k_l\ell_i}\notag{}\\
             &-\hefac\text{sgn}(\ell_i-\ell_s)\mathbf d_i^*\cdot(\hat e_z\cross\bm\mu_s)\me^{\iu k_z\abs{\ell_i-\ell_s}}\beta_m
  -\efac\mathbf d_i^*\cdot\mathbf d_m\me^{\iu k_z\abs{\ell_i-\ell_s}}\beta_j,\\
  \dv{\beta_s}{t}&=(\iu\Delta_s-\gamma_s/2)\beta_s
                  -\efac\sum_{i}\mathbf d_s^*\cdot\mathbf d_i\me^{\iu k_z\abs{\ell_i-\ell_j}}\beta_i
                  -\iu\mathbf d_s^*\cdot \mathbf E\ts{in}\me^{\iu k_l\ell_s}
                  -\iu\bm \mu_i^*\cdot \mathbf B\ts{in}\me^{\iu k_l\ell_s}\notag{}\\
  &-\hefac\sum_i\text{sgn}(\ell_i-\ell_s)\bm\mu_s^*\cdot(\hat e_z\cross\mathbf d_i)\me^{\iu k_z\abs{\ell_i-\ell_s}}\beta_i,
\end{align}
\end{subequations}
where $\beta_i$ are the coherences of the metasurfaces and $\beta_s$ is the coherence of the molecular transition. $\ell_i$ and $\ell_s$ denote the positions of the metasurfaces and the molecular scatterer along the $z$ axis. This neglects the coupling between the molecule and all other surface modes with different momenta. This can be taken into account by renormalizing the decay rate $\gamma_s=\efac d_s^2+\hfac \mu_s^2$ which will however only affect the total transmission which is of no particular interest to us.

For the proposed setup of four metasurface layers with dipole moments $(d\hat e_x,d\hat e_y,d\hat e_x,d\hat e_y)$ at positions $(0,\ell_m,\ell,\ell+\ell_m)$ with $\ell_m =\lambda/4$, it is thus clear that the superpositions $\beta^x_L+\xi \beta^y_L$ and $\beta^x_R+\xi \beta^y_R$ couple to a specific polarization component. For left or right-handed illumination, we thus consider the equations of motion only in this subspace

\begin{subequations}
\begin{align}
  \label{eq:four_surfaces_projected}
    \dv{\beta^\xi_L}{t}&=(\iu\Delta-\widetilde\Gamma/2)\beta^\xi_L-\frac{\widetilde\Gamma}{2}\me^{\iu k_z\ell} \beta_R^\xi-\frac{\sqrt{\widetilde\Gamma}}{2}\me^{\iu k_z\ell_s}\hat e_{-\xi}\cdot\left[\sqrt{\gamma^e_s}\hat d_s+\iu\xi\sqrt{\gamma^m_s}\hat\mu_s\right]\beta_s-\iu\sqrt{\widetilde\Gamma}E\ts{in},\\
    \dv{\beta^\xi_R}{t}&=(\iu\Delta-\widetilde\Gamma/2)\beta^\xi_R-\frac{\widetilde\Gamma}{2}\me^{\iu k_z\ell}\beta^\xi_L-\frac{\sqrt{\widetilde\Gamma}}{2}\me^{\iu k_z(\ell-\ell_a)}\hat e_\xi\cdot\left[\sqrt{\gamma^e_s}\hat d_s+\iu\xi\sqrt{\gamma^m_s}\hat \mu_m\right]\beta_s-\iu\sqrt{\widetilde\Gamma}E\ts{in}\me^{\iu k_z\ell},\\
    \dv{\beta_s}{t}&=[\iu\Delta_s-(\gamma^m_s+\gamma^e_s)/2]\beta_s-\frac{\sqrt{\widetilde\Gamma}}{2}\me^{\iu k_z\ell_s}\left[\sqrt{\gamma^e_s}\hat d_s^*-\iu\xi\sqrt{\gamma^m_s}\hat\mu_s^*\right]\cdot\hat e_{-\xi}\beta^\xi_L\nonumber\\
    &-\frac{\sqrt{\widetilde\Gamma}}{2}\me^{\iu k_z(\ell-\ell_s)}\left[\sqrt{\gamma^e_s}\hat d_s^*-\iu\xi\sqrt{\gamma^m_s}\hat\mu_s^*\right]\cdot\hat e_\xi \beta_R^\xi.
\end{align}
\end{subequations}
Here $\xi=\pm$ indicates the incoming polarization and we have split the decay rate of the molecular transition into a magnetic and electric part $\gamma_s=\gamma^m_s+\gamma^e_s$. The vectors $\hat e_\pm$ are the circular basis vectors. Assuming steady state for $\beta_s$ in the dispersive regime $\Delta_s\gg\gamma_s$, one obtains two terms for the left and right side. One term that renormalizes the coupling between the metasurface layers and another term which renormalizes the metasurface linewidth and frequency. The second term is of the form
\begin{equation}
  \label{eq:cavity_renorm_1}
  -\iu\frac{\widetilde\Gamma}{4\Delta_s}\me^{2\iu k_z\ell_s}\hat e_{-\xi}\cdot\left[\sqrt{\gamma^e_s}\hat d_m+\iu\xi\sqrt{\gamma^m_s}\hat\mu_s\right]\left[\sqrt{\gamma^e_s}\hat d_s^*-\iu\xi\sqrt{\gamma^m_s}\hat\mu_s^*\right]\cdot\hat e_{-\xi}
\end{equation}
for the left and
\begin{equation}
  \label{eq:cavity_renorm_2}
  -\iu\frac{\widetilde\Gamma}{4\Delta_s}\me^{2\iu k_z(\ell-\ell_s)}\hat e_{\xi}\cdot\left[\sqrt{\gamma^e_s}\hat d_s+\iu\xi\sqrt{\gamma^m_s}\hat\mu_s\right]\left[\sqrt{\gamma^e_s}\hat d_s^*-\iu\xi\sqrt{\gamma^m_s}\hat\mu_s^*\right]\cdot\hat e_{\xi}
\end{equation}
for the right.

It is well known that a rotational average of the terms in brackets yields the same result for both terms
\begin{equation}
  \label{eq:cavity_renorm_rotational_average}
  \expval{\hat e_{\pm\xi}\cdot\left[\sqrt{\gamma^e_s}\hat d_s+\iu\xi\sqrt{\gamma^m_s}\hat\mu_s\right]\left[\sqrt{\gamma^e_s}\hat d_s^*-\iu\xi\sqrt{\gamma^m_s}\hat\mu_s^*\right]\cdot\hat e_{\pm\xi}}_{\text{rot}}
  =\left[\gamma_s+2\xi\sqrt{\gamma^m_s\gamma_s^e}\Im(\hat d_s^*\cdot\hat\mu_s)\right],
\end{equation}
where $\iu\sqrt{\gamma^m_s\gamma_s^e}\Im(\hat d_s^*\cdot\hat\mu_s)$ is usually identified as the rotary strength of the molecular transition which is nonzero for chiral molecules and zero for achiral molecules. The limit of a perfect chiral scatterer for $\hat d_s\in\mathbb{R}$ and $\hat\mu_s\in\iu\mathbb{R}$ is when the two are aligned and have the same magnitude so that $\sqrt{\gamma^m_s\gamma_s^e}\Im(\hat d_s^*\cdot\hat\mu_s)=\pm \gamma_s$. This implies that the scatterer only couples to one of the polarization components.

\section{Derivation of phase uncertainty for homodyne detection}\label{sec:AppendixI}

\begin{table}[h!]
  \centering
  \begin{tabular}{||l|l|l||}\toprule
    $M_- (t, T)$ & intensity difference operator &\\
    $m_-$ & detected intensity difference  & \\
    $F$ & intensity of signal beam ($\#$photons/time)  & unit [Hz] \\
    $F_\mathrm{LO}$ & intensity of LO beam ($\#$photons/time)  & unit [Hz] \\
    $T$ & detector integration time & \\
    $\eta_Q$ & quantum efficiency of detector  & $0\leq\eta_Q\leq 1$\\
    $\theta$ & phase of signal beam & \\
    $\theta_\mathrm{LO}$ & phase of LO  & \\
    \bottomrule
  \end{tabular}
  \caption{Definitions for section S9.}
  \label{tab:defs9}
\end{table}

We consider balanced homodyne detection and we follow closely section 6.11 of \cite{LoudonBook}. First, we define the intensity difference operator
\begin{align}\label{a200}
  M_-(t,T) = \mi \int_{t}^{t+T} \di t'  \Bigl[t_c^* \ha^\dagger(t') \ha_\mathrm{LO}(t') - t_c \ha^\dagger_\mathrm{LO}(t') \ha(t') \Bigr],
\end{align}
where $T$ is the detector integration time ($T$ and $t$ are not to be confused with the cavity transmission function $t_c$), $\ha(t)$ is given by
\begin{align}\label{a90}
  \ha(t) = \frac{1}{\sqrt{2 \pi}} \int_{-\infty}^\infty \di \omega\,  \ha(\omega)  \me^{-\mi \omega t}.
\end{align}
and $\ha_\mathrm{LO}(t)$ is the annihilation operator of the local oscillator field, still defined, mutatis mutandis, by Eq.~\eqref{a90}. The signal and the LO are prepared in the \emph{monochromatic} coherent states $\ket{\alpha}$ and $\ket{\alpha_\mathrm{LO}}$, respectively, such that
\begin{align}\label{a210}
  \ha(t) \ket{\alpha} = \alpha(t) \ket{\alpha}, \qquad \text{and} \qquad \ha_\mathrm{LO}(t) \ket{\alpha_\mathrm{LO}} = \alpha_\mathrm{LO}(t) \ket{\alpha_\mathrm{LO}},
\end{align}
where
\begin{align}\label{a220}
  \alpha(t) = \sqrt{F}  \exp (- \mi \omega_0 t + \mi \theta), \qquad \text{and} \qquad  \alpha_\mathrm{LO}(t) = \sqrt{F_\mathrm{LO}}  \exp (- \mi \omega_0 t +\mi \theta_\mathrm{LO}).
\end{align}
By definition $F$ and $F_\mathrm{LO}$ are the number of photons per unit of time of the signal and the LO beams, respectively. In a typical experimental setup $F_\mathrm{LO} \gg F$.

The detected intensity difference $m_-$ is calculated as
\begin{align}\label{a230}
  m_- = \eta_Q \expe{M_-(t,T)},
\end{align}
where $\eta_Q$ is the quantum efficiency of the detectors and here and hereafter $\ave{O} = \mean{\alpha,\alpha_\mathrm{LO}}{O}{\alpha,\alpha_\mathrm{LO}}$. A straightforward calculation gives
\begin{align}\label{a240}
  \expe{M_-(t,T)} = 2\sqrt{F  F_\mathrm{LO}}\abs{t_c}\int_{t}^{t+T}\diff{t'}\sin \left( \theta - \theta_\mathrm{LO} + \varphi(t') \right),
\end{align}
where we have defined $t_c = \abs{t_c} \exp(\mi \varphi(t))$. Since we consider the case of a particle passing through the cavity, we keep the phase time-dependent but assume the transmission is not affected significantly by the passage.

The difference photocount variance $(\Delta m_-)^2$  is given by
\begin{align}\label{a250}
  (\Delta m_-)^2 = \eta_Q^2 \left\{ \expe{\left[ M_-(t,T) \right]^2} - \expe{M_-(t,T)}^2 \right\} + \eta_Q(1-\eta_Q)
  \int_{t}^{t+T} \di t'  \expe{  \ha_\mathrm{LO}^\dagger(t') \ha_\mathrm{LO}(t') - \abs{t_c}^2 \ha^\dagger(t') \ha(t') },
\end{align}
where
\begin{align}\label{a260}
  \expe{\left[ M_-(t,T) \right]^2} = \abs{t_c}^2 \left[T \left( F + F_\mathrm{LO} \right) + 4 F F_\mathrm{LO}  \left(\int_{t}^{t+T}\diff{t'}\sin \left( \theta - \theta_\mathrm{LO} + \varphi(t') \right)\right)^2 \right].
\end{align}
Gathering all the pieces together, we eventually find
\begin{align}\label{a270}
  (\Delta m_-)^2 = \eta_Q  T \Bigl[ \eta_Q  \abs{t_c}^2 \left( F + F_\mathrm{LO} \right)+ \left( 1-\eta_Q \right) \left( \abs{t_c}^2 F + F_\mathrm{LO} \right) \Bigr].
\end{align}
Small changes in the transmission function of the Fabry-P\'erot cavity due to the passage of the particle produce small changes in the mean photocounts at the two detectors in the homodyne setup. The limiting resolution is determined by the intrinsic uncertainty in the homodyne detection, quantified by the difference photocount variance $(\Delta m_-)^2 $. The uncertainty $\Delta \varphi$ in the transmission function caused by the homodyne uncertainty is obtained from
\begin{align}\label{a280}
  \Delta \varphi = \frac{\sqrt{(\Delta m_-)^2}}{\displaystyle \abs{\frac{\di m_-}{\di \varphi}}}.
\end{align}
From Eqs.~\eqref{a230} and \eqref{a240} we have
\begin{align}\label{a290}
  \abs{\frac{\di m_-}{\di \varphi}} = 2  \eta_Q  T  \sqrt{F  F_\mathrm{LO}}   \abs{t_c} \abs{\int_{t}^{t+T}\diff{t'}\cos \left( \theta - \theta_\mathrm{LO} + \varphi(t') \right)}.
\end{align}
For a signal and a local oscillator matched with one of the cavity resonances $\omega_m = 2 \pi (m-1/2)/\tR$ we have $\abs{t_c}=1$ and $\theta - \theta_\mathrm{LO} + \varphi(t)=\delta\varphi(t)$, so that
\begin{align}\label{a300}
  (\Delta m_-)^2_\text{res} = \eta_Q  T   \left( F + F_\mathrm{LO} \right), \qquad \text{and} \qquad \abs{\frac{\di m_-}{\di \varphi}}_\text{res} = 2  \eta_Q  T  \sqrt{F  F_\mathrm{LO}}\abs{\int_{t}^{t+T}\diff{t'} \cos \left( \delta\varphi(t')\right)}.
\end{align}
Substituting Eq.~\eqref{a300} into \eqref{a280} we obtain
\begin{align}\label{a310}
  \Delta \varphi = \frac{\sqrt{F + F_\mathrm{LO}}}{2  \sqrt{\eta_Q  T}  \sqrt{F  F_\mathrm{LO}} \abs{\int_{t}^{t+T}\diff{t'} \cos \left( \delta\varphi(t')\right)}} \approx \frac{1}{2  (\eta_Q T  F)^{1/2} \abs{\int_{t}^{t+T}\diff{t'} \cos \left( \delta\varphi(t')\right)}},
\end{align}
where the last approximate equality holds when $F_\mathrm{LO} \gg F$. Similarly, in this regime

\begin{equation}
  \label{eq:approximate_error}
  \frac{\sqrt{(\Delta m_-)^2_\text{res}}}{\sqrt{F_\mathrm{LO}}}= \sqrt{\eta_Q T}.  
\end{equation}

\end{document}